\documentclass[jkps,twocolumn,showpacs,floatfix]{revtex4}
\usepackage[dvips]{graphicx}
\begin{document}
\title{Empirical Formula for the Excitation Energies\\ of the First
$2^+$ and $3^-$ States in Even-even Nuclei}
\author{Eunja \surname{Ha}}
\author{Dongwoo \surname{Cha}}
\email{dcha@inha.ac.kr}
\thanks{Fax: +82-32-866-2452}
\affiliation{Department of Physics, Inha University, Incheon
402-751, Korea}
\date{December 21, 2006}

\begin{abstract}
We report empirical findings that a simple formula in terms of the
mass number $A$, the valence proton number $N_p$, and the valence
neutron number $N_n$ can describe the essential trends of the
excitation energies $E_x$ of the first $2^+$ and $3^-$ states in
even-even nuclei throughout the periodic table. The formula reads as
$E_x = \alpha A^{-\gamma} + \beta \left[ \exp (- \lambda N_p ) +
\exp (- \lambda N_n ) \right] $. The parameter $\gamma$ in the first
term is determined from the mass number $A$ dependence of the bottom
contour line of the excitation energy systematics. The other three
parameters $\alpha$, $\beta$, and $\lambda$ are fitted by minimizing
the $\chi^2$ value between logarithms of the measured and the
calculated excitation energies. Our results suggest that the single
large-$j$ shell simulation can be applied to the excitation energies
of the first $2^+$ and $3^-$ states in even-even nuclei.
\end{abstract}

\pacs{21.10.Re, 23.20.Lv}

\maketitle

The exhaustive compilation of the $B\,(E2)$ values between the $0^+$
ground state and the first $2^+$ state in even-even nuclei by Raman
{\it et al}. has provided a rare opportunity to make a systematic
study of the relevant nuclear properties throughout the periodic
table \cite{Raman-a,Raman-b}. For example, it was seen that
$B\,(E2)$ values, when plotted in terms of the atomic number $Z$,
showed a mid-shell bump within a major shell consisting of several
single -particle levels. It was also shown that this bump could be
nicely explained by an idea of the single large-$j$ shell simulation
\cite{Raman-c,Raman-d}. Within this simulation, the nuclei
corresponding to the $j_z$ sub-states of large angular momentum $j$
are described in terms of the valence nucleon numbers $N_p$ and
$N_n$, which are then adopted to parameterize various quantities
such as the $B(E2)$ values and the nuclear quadrupole moments. The
valence proton (neutron) number $N_p\,(N_n )$ is defined as the
number of proton (neutron) particles below mid-shell or the number
of proton (neutron) holes past mid-shell within the given major
shell. Hamamoto, who introduced the concept of the valence nucleon
numbers for the first time more than four decades ago, has shown
that the square root of the ratio of the measured and the
single-particle $B(E2)$ values, $ \left[ B(E2)_{\rm exp} /
B(E2)_{\rm sp} \right]^{1/2} $, is roughly proportional to the
product $N_p N_n $ \cite{Hamamoto}. Casten has extended the idea and
shown that if one parameterizes the collective variables or
operators in terms of the product $N_p N_n$ instead of the usual
mass number $A$, the neutron number $N$, or the proton number $Z$,
then one gets a substantial reduction in the number of parameters
without serious loss of accuracy \cite{Casten,Casten-a}.

\begin{figure*}
\centering
\includegraphics[width=15cm,angle=0]{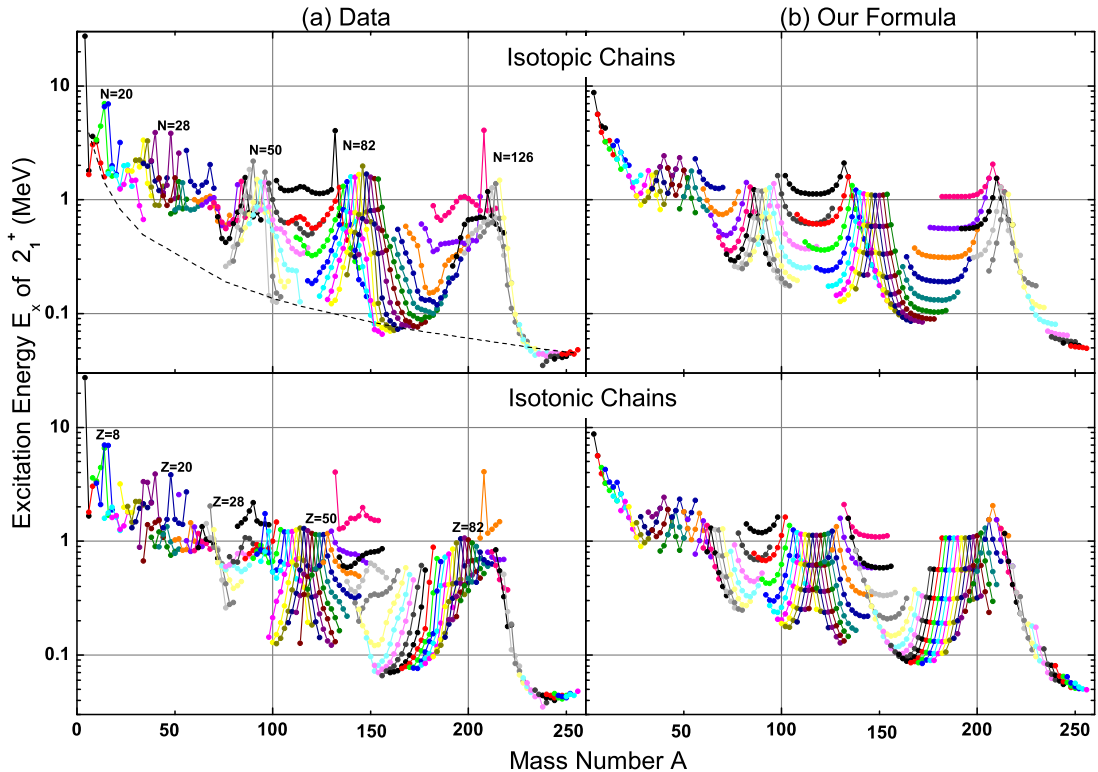}
\caption{Excitation energies of the first $2^+$ states in
even-even nuclei. The solid lines in the upper (lower) panels represent
isotopic (isotonic) chains. The left panels denoted by (a) show the
data quoted from Ref. 2, and the right panels denoted
by (b) show the results obtained by our formula, Eq.\,(\ref{E}). (In
the electronic version, the color code for an isotopic (isotonic) chain
in part (a) is the same as the color code for the corresponding
isotopic (isotonic) chain in part (b).)} \label{fig-1}
\end{figure*}

In this brief report, we want to illustrate that the valence nucleon
numbers can be employed in describing not only the $B(E2)$ values as
shown in the previous publications by other authors
\cite{Hamamoto,Casten} but also the excitation energies themselves
of the first $2^+$ states in the even-even nuclei. In addition, we
will also show that a single formula in terms of the valence
nucleon numbers reproduces the essential trends of the first $2^+$
excitation energies throughout the periodic table. This fact
presents a noticeable contrast to former studies in which the
parametrization by the valence nucleon numbers $N_p$ and $N_n$ was
performed for nuclei that belonged to one major shell only.

Evidence in favor of the single large-$j$ shell structure in the
excitation energies $E_x (2_1^+ )$ of the first $2^+$ states can be
seen from Fig.\,\ref{fig-1}(a), which shows the data quoted from
Ref. 2 where the best-known values of $E_x (2_1^+ )$ have been
compiled for even-even nuclei. First, consider the solid lines in
the upper panel of Fig.\,\ref{fig-1}(a), which connect the
excitation energies of the nuclei that belong to the same isotopic
chain. Let us focus on the V-shaped portion that can be easily
identified from each one of isotopic chains in Fig.\,\ref{fig-1}(a).
Those nuclei which form a Vshape actually constitute one of the
neutron major shells (with the proton number fixed) in the single
large-$j$ shell simulation. The Vshape itself tells us that, within
a major shell, the excitation energy $E_x (2_1^+)$ is minimum in the
mid-shell nucleus, which corresponds to the bottom of the Vshape,
and
 that $E_x (2_1^+)$ becomes larger either when the neutron number
increases or decreases from the mid-shell. Furthermore, as indicated
by the numbers in the upper panel of Fig.\,\ref{fig-1}(a), the neutron
number $N$ of the nucleus at the top of Vshapes is one of
magic numbers $8$, $20$, $28$, $50$, $82$, and $126$ that form the
boundary between adjacent major shells within the model of the
single large-$j$ shell simulation \cite{Raman-d}. Now, turn to the
lower panel of Fig.\,\ref{fig-1}(a). The data points in the lower
panel are exactly the same as those in the upper panel, but the
solid lines in the lower panel of Fig.\,\ref{fig-1}(a) are obtained
by connecting the excitation energies of the nuclei that belong to
the same isotonic chain with different proton numbers. From the
isotonic chains in the lower panel of Fig.\,\ref{fig-1}(a), we can
observe precisely the same kind of evidence for the single large-$j$
proton shell structure (with the neutron number fixed) as for the
single large-$j$ neutron shell structure in the upper panel of
Fig.\,\ref{fig-1}(a).

As an effort to describe the systematic behavior of $E_x (2_1^+ )$
by using a formula that is as simple as possible, we try the following
equation:
\begin{equation} \label{E}
E_x = \alpha A^{-\gamma} + \beta \left[ \exp ( - \lambda N_p ) +
\exp ( - \lambda N_n ) \right],
\end{equation}
where $\alpha$, $\beta$, $\gamma$, and $\lambda$ are our free
parameters to be fitted from the data. The first term of
Eq.\,(\ref{E}) represents the overall dependence of the excitation
energy on the mass number $A$, and the last two terms account for
variations of excitation energies, which form the Vshape within a
major shell in Fig.\,\ref{fig-1}. We divide our fitting procedure
into two steps. First, we construct the bottom contour line by
connecting the lowest points among the adjacent V's in the upper
panel of Fig.\,\ref{fig-1}(a) and fit it by only the first term, $
\alpha A^{-\gamma} $, of Eq.\,({\ref{E}) to determine the two
parameters $\alpha$ and $\gamma$. However, we take only the value of
$\gamma$ from the first step, and let $\alpha$, $\beta$, and
$\lambda$ be subject to a variation in minimizing the ${\chi}^2$
value given by
\begin{equation} \label{Chi}
\chi^2 = { 1 \over {N_0}} \sum_{i=1}^{N_0} \left\{ \ln \left[
E_x^{\rm cal} (i) \right] - \ln \left[ E_x^{\rm exp} (i) \right]
\right\}^2
\end{equation}
between logarithms of all the measured excitation energies $E_x^{\rm
exp}$ in Fig.\,\ref{fig-1}(a) and the $E_x^{\rm cal}$ calculated by
using Eq.\,(\ref{E}), where $N_0$ is the number of total data points
considered. The adopted values of $\alpha$, $\beta$, $\gamma$, and
$\lambda$ in Eq.\,(\ref{E}) for the excitation energies of the first
$2^+$ states are listed in the first row of Table \ref{tab-1}. Our
adopted bottom contour line, $E_x = \alpha A^{-\gamma}$, is shown as
the dashed curve in the upper panel of Fig.\,{\ref{fig-1}(a).

\begin{table}[b]
\begin{center}
\caption{Values adopted for the three parameters in Eq.\,(\ref{E})
for the excitation energies of the two multipole states $2_1^+$ and
$3_1^-$.}
\begin{tabular}{ccccc}
\hline\hline
Multipole&~~~$\alpha$(MeV)~~~&~~~$\beta$(MeV)~~~&~~~~~~$\gamma$~~~~~~&~~~~~~$\lambda$~~~~~~\\
\hline
$2_1^+$&34.9&1.00&1.19&0.36\\
$3_1^-$&62.8&0.90&0.79&0.24\\
\hline\hline
\end{tabular}
\label{tab-1}
\end{center}
\end{table}

In Fig.\,\ref{fig-1}(b), the excitation energies of the
first $2^+$ states in even-even nuclei calculated with our formula,
Eq.\,(\ref{E}), with the parameter set given in Table \ref{tab-1}
are plotted. The upper and the lower panels of Fig.\,\ref{fig-1}(b) show
the same results, but the curves in the upper panel connect isotopic
chains while those in the lower panel connect isotonic chains. One
can immediately find a very close resemblance between the curves in
Figs.\,\ref{fig-1}(a) and (b), the data and our results,
respectively. (In the electronic version, the color code for an
isotopic (isotonic) chain in Fig.\,\ref{fig-1}(a) is the same as the
color code for the corresponding isotopic (isotonic) chain in
Fig.\,\ref{fig-1}(b).) It is quite remarkable to note that a simple
formula such as Eq.\,(\ref{E}) can reproduce the data not only
qualitatively but also quantitatively to some extent. Especially, the
characteristic shape of the single large-$j$ neutron shell structure
in isotopic chains and that of the single large-$j$ proton
shell structure in isotonic chains is nicely reproduced by our
formula.

\begin{figure}[b]
\centering
\includegraphics[width=8.0cm,angle=0]{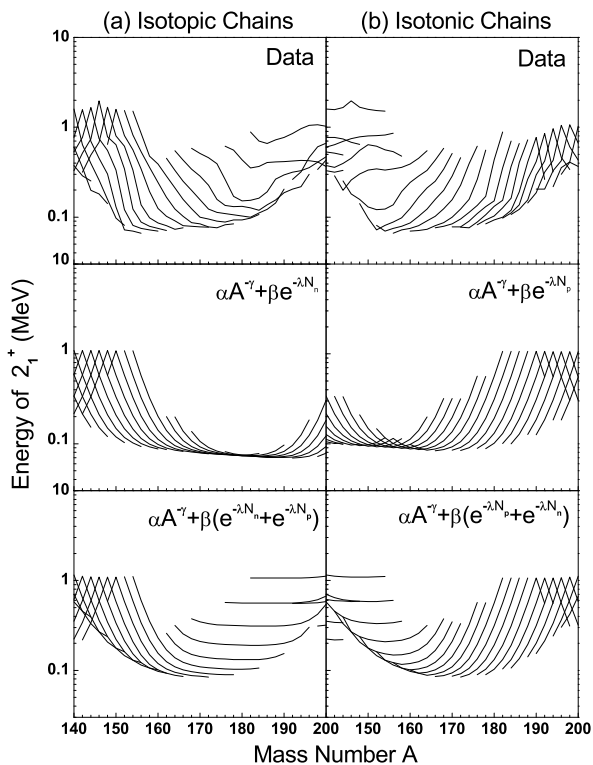}
\caption{Excitation energies $E_x (2_1^+ )$ for nuclei between
$140 \le A \le 200$. The solid lines in (a) and (b) represent the
isotopic and the isotonic chains, respectively. The top, central, and
bottom panels show the data, the excitation energies calculated with
only the first two terms of Eq.\,(\ref{E}), and those calculated
with all three terms of Eq.\,(\ref{E}), respectively.}
\label{fig-2}
\end{figure}

The role played by the two exponential terms of Eq.\,(\ref{E}) in
reproducing the data can be seen in Fig.\,\ref{fig-2} where the
excitation energies $E_x (2_1^+ )$ for nuclei between $140 \le A \le
200$ are displayed. The solid lines of the left part (a) of
Fig.\,\ref{fig-2} are obtained by connecting the excitation energies
that belong to isotopic chains while those of the right part (b) are
obtained by connecting those that belong to isotonic chains. Let us
first compare the central two panels and then the bottom two panels
with the data in the top two panels. The solid lines of the left
central panel are drawn by taking only the first two terms of
Eq.\,(\ref{E}), which show $ \alpha A^{-\gamma} + \beta \exp ( -
\lambda N_n )$, and those of the right central panel are drawn by
taking two similar terms, $\alpha A^{-\gamma} + \beta \exp (-
\lambda N_p )$. We can observe that while the lines at the left
(right) side of the left (right) central panel explain the data
shown in the corresponding top panel quite reasonably, the same
lines at the right (left) side of the left (right) central panel
cannot reproduce the main features of the data, which show a
monotonic increase in the excitation energy for increasing
(decreasing) mass number $A$. Now, by comparing the top two panels
with the corresponding bottom ones, which are obtained by adding the
term $ \beta \exp (- \lambda N_p )$ in the case of the left central
panel or the term $ \beta \exp (- \lambda N_n )$ in the case of the
right central panel, we find that the overall main features of the
data are fairly well reproduced.

\begin{figure}[b]
\centering
\includegraphics[width=8.0cm,angle=0]{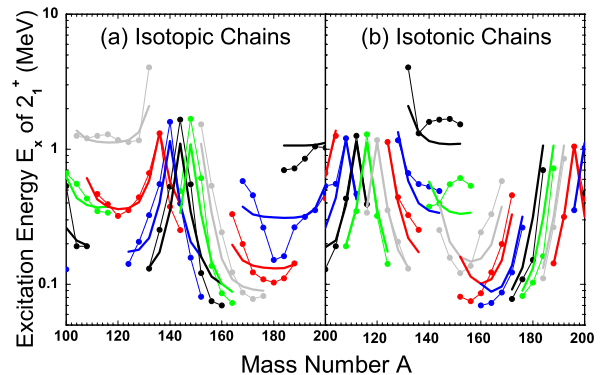}%
\caption{Measured and calculated excitation energies $E_x
(2_1^+ )$ for nuclei between $100 \le A \le 200$. The solid lines in
(a) and (b) represent the isotopic and the isotonic chains, respectively.
The calculated excitation energies are represented by thick solid
curves while the measured ones are depicted by solid circles and
are connected by thin solid lines.} \label{fig-3}
\end{figure}

One can observe from Fig.\,\ref{fig-1} that the measured excitation
energies $E_x (2_1^+ )$ for nuclei roughly between $100 \le A \le
200$ are explained best by our formula given by Eq.\,(\ref{E}), even
quantitatively to some extent. In order to examine them in more
detail, we extract the corresponding part from Fig.\,\ref{fig-1},
and out of it we create Fig.\,\ref{fig-3} where the calculated
excitation energies $E_x (2_1^+ )$ and the data are plotted in the
same panel. In Fig.\,\ref{fig-3}(a), the calculated excitation
energies of the isotopic chains are represented by thick solid
curves while the measured ones of the corresponding isotopic chains
are depicted by solid circles connected by thin solid lines. Figure
\ref{fig-3}(b) is drawn from exactly the same data points used in
Fig.\,\ref{fig-3}(a), but the points are connected for isotonic
chains. However, three quarters of the data points that appear in
Fig.\,\ref{fig-1} were excluded in making Fig.\,\ref{fig-3} to avoid
too much complexity within a graph. It is noticeable in
Fig.\,\ref{fig-3} that our formula, which is able to account for the
essential trends in the systematic behavior of the excitation
energies $E_x (2_1^+ )$ observed in Fig.\,\ref{fig-1} very
successfully, has its shortcomings in explaining either the detailed
shape or actual height of the measured isotopic or isotonic chains.

\begin{figure}[t]
\centering
\includegraphics[width=8.0cm,angle=0]{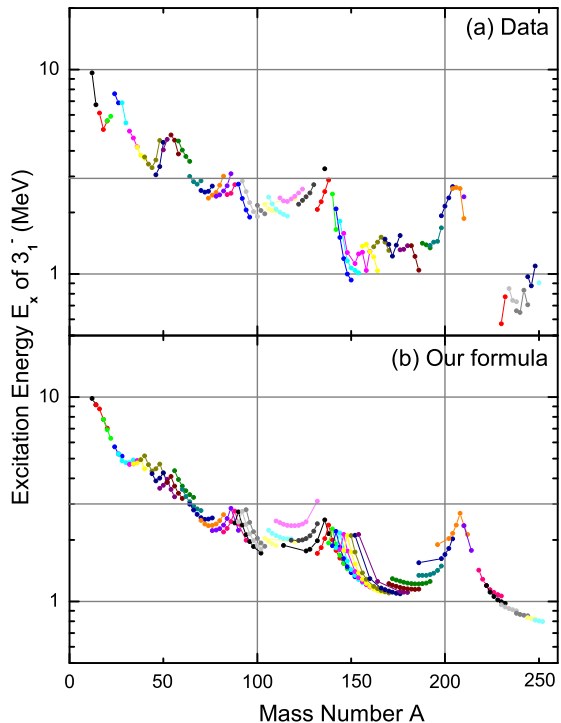}%
\caption{Excitation energies of the first $3^-$ states in
even-even nuclei. The solid lines connecting the data points represent
isotopic chains. Part (a) shows the data quoted from Ref. 8,
and part (b) shows the results obtained by using our
formula, Eq.\,(\ref{E}).} \label{fig-4}
\end{figure}

Encouraged by uncovering the fact that the valence nucleon numbers
can be employed successfully in explaining the excitation energies
of the first $2^+$ states in even-even nuclei throughout the
periodic table, we decided to apply the above same procedure to the
first $3^-$ states in even-even nuclei. Fortunately, a recently
updated compilation of the excitation energies $E_x (3_1^- )$ and
the $B(E3)$ values between the $0^+$ ground state and the first
$3^-$ state has been already published by Kib\'{e}di and Spear
\cite{Kibedi}.

The results for $E_x (3_1^- )$ are presented in Fig.\,\ref{fig-4}.
The measured excitation energies shown in Fig.\,\ref{fig-4}(a) are
quoted from Ref. 8, and the calculated ones plotted in
Fig.\,\ref{fig-4}(b) are obtained by using Eq.\,(\ref{E}). The solid
lines connecting the data points in Figs.\,\ref{fig-4} (a) and (b)
represent isotopic chains. Additional data points, which do not have
measured counterparts in the part (a), are included in the part (b)
for illustrative purposes. The four parameters, $\alpha$, $\beta$,
$\gamma$, and $\lambda$, in Eq.\,(\ref{E}) for $E_x (3_1^- )$ are
determined by exactly the same fitting procedure as for $E_x (2_1^+
)$, and the final adopted values are listed in the second row of
Table \ref{tab-1}. One can find a very close resemblance between (a)
the data and (b) our results in Fig.\,\ref{fig-4} and conclude that
the characteristic shape of the single large-$j$ shell structure is
present in the excitation energies of the first $3^-$ states as well
as of the first $2^+$ states, in even-even nuclei. The excitation
energies estimated by using our simple formula may be of assistance
in the various nuclear structure studies \cite{Lee-a,Lee-b}.

In summary, we have presented a simple formula that can describe the
systematic behavior of excitation energies of the first $2^+$ and
$3^-$ states in even-even nuclei. Our formula, given by Eq.
(\ref{E}), is composed of only three terms that depend on the mass
number $A$, the valence proton number $N_p$, and the valence neutron
number $N_n$, respectively. The first term of Eq. (\ref{E})
represents the overall dependence of the excitation energy on the
mass number $A$ while the second (third) term reflects the single
large-$j$ proton (neutron) shell structure found in each isotonic
(isotopic) chain. We find that Eq. (\ref{E}) can reproduce the
essential trends of the measured excitation energies of the first
$2^+$ and $3^-$ states, which were compiled extensively in Ref. 2
and Ref. 8, respectively. Furthermore, a preliminary study indicates
that Eq. (\ref{E}) can still be applied to the first $4^+$
excitation energies and to the second $2^+$ excitation energies in
even-even nuclei \cite{Kim}. Therefore, it would be most interesting
if one can clarify the physics underlying the above observations.

\begin{acknowledgments}
This work was supported by a Inha University research grant.
\end{acknowledgments}

\end{document}